# Sentiment analysis of twitter data

Hamid Bagheri
Computer Science Department
Iowa State University
hbagheri@iastate.edu

Md Johirul Islam
Computer Science Department
Iowa State University
mislam@iastate.edu

*Abstract*—Social networks are the main resources to gather information about people's opinion and sentiments towards different topics as they spend hours daily on social medias and share their opinion. In this technical paper, we show the application of sentimental analysis and how to connect to Twitter and run sentimental analysis queries. We run experiments on different queries from politics to humanity and show the interesting results. We realized that the neutral sentiment for tweets are significantly high which clearly shows the limitations of the current works.

I. INTRODUCTION

Opinion and sentimanetal mining is an important resarch areas because due to the huge number of daily posts on social networs, extracting people's opionin is a challenging task. About 90 percent of today's data has been provided during the last two years and getting insight into this large scale data is not trivial [17, 18].

Sentimental analysis has many applications for different domains for example in businesses to get feedbacks for products by which companies can learn users's feedback and reviews on social medias.

Opinion and sentimental mining has been well studied in this reference and all different approaches and research fields have been discussed [10]. There are also some works have been done on Facebook [19-23] sentimental analysis however in this paper we mostly focus on the Twitter sentimental analysis.

For a larger texts one solution could be understand the text, summarize it and give weight to it whether it is positive, negative or neutral. Two fundamental approaches to extract text summarization are an extractive and abstractive method. In the extractive method, words and word phrases are extracted from the original text to generate a summary. In an abstractive method, tries to learn an internal language representation and then generates summary that is more similar to the summary done by human.

Text understanding is a significant problem to solve. Some machine learning techniques, including various supervised and unsupervised algorithms, are being utilized. There are different approaches to generate summary. One approach could be rank the importance of sentences within the text and then generate summary for the text based on the importance numbers. There is another approach called end-to-end generative models. In some domain like image recognition, speech recognition, language translation, and question-answering, the end-to-end method performs better.

Some works have used an ontology to understand the text [1]. In the phrase level, sentimental analysis system should be able to recognize the polarity of the phrase which is discussed by Wilson, et.al [9]. Tree kernel and feature based model have been applied for sentimental analysis in twitter by Agarwal and et.al [11]. SemEval-2017 [12] also shows the seven years of sentimental analysis in twitter tasks. Since tweets in Twitter is a specific text not like a normal text there are some works that address this issue like the work for short informal texts [13]. Sentimental analysis has many applications in news [14].

In this paper, we will discuss social network analysis and the importance of it, then we discuss Twitter as a rich resource for sentimental analysis. In the following sections, we show the high-level abstract of our implementation. We will show some queries on different topics and show the polarity of tweets.

## II. SOCIAL NETWORK ANALYSIS

Social netowrk analysis is the study of people's interactions and communications on different topics and nowadays it has received more attention. Million of people give their opinion of different topics on a daily basis on social medias like Facebook and Twitter. It has many applications in different areas of research from social science to business [3].

Twitter nowadays is one of the popular social media which according to the statistain [4] currently has over 300 millions accounts. Twitter is the rich source to learn about people's opion and sentimental analysis [2]. For each tweet it is important to determine the sentiment of the tweet whether is it positive, negative, or neutral.

Another challenge with twitter is only 140 characters is the limitaiton of each tweet which cause people to use phrases and works which are not in language processing. Recently twitter has extended the text limitations to 280 characters per each tweet.

## III. TWITTER SENTIMENTAL ANALYSIS

Social networks is a rich platform to learn about people's opinion and sentiment regarding different topics as they can communicate and share their opinion actively on social medias including Facebook and Twitter. There are different opinion-oriented information gathering systems wich aim to extract people's opinion regarding different topics. The sentiment-aware systems these days have many applications from business to social siences.

Since social networks, especially Twitter, contains small texts and people may use different words and abbreviations which are difficult to extract their sentiment by current Natural Language processing sysntems easily, therefore some researchers have used deep learning and machine learning techniques to extract and mine the polarity of the text [15]. Some of the top abbreviations are FB for facebook, B4 for before, OMG for oh my god and so on. Therefore sentimental analysis for short texts like Twitter's posts is challengeing [8].

## IV. DESIGN AND IMPLEMENTATION

This technical paper reports the implementation of the Twitter sentiment analysis, by utilizing the APIs provided by Twitter itself.

There are great works and tools focusing on text mining on social networks. In this projecct the welth of available libararies has been used.

The approach to extract sentiment from tweets is as follows:

1. Start with downloading and caching the sentiment dictionary

2. Download twitter testing data sets, input it in to the program.

3. Clean the tweets by removing the stop words.

4. Tokenize each word in the dataset and feed in to the program.

5. For each word, compare it with positive sentiments and negative sentiments word in the dictionary. Then increment positive count or negative count.

6. Finally, based on the positive count and negative count, we can get result percentage about sentiment to decide the polarity.

Researchers have done different sentimental analysis on Twitter for different purposes for example the work designed by Wang, *et.al* [5] is a real-time twitter sentimental analysis of the presidential elections.

Figure 1 shows the sentimental analysis algorithm at the high level. As it can be seen in the algorithm, we have different procedures to connect the twitter API, fetch the tweets, tweet cleaning or remove stop words, classify tweets which means get the polarity of the tweet, and finally return the results.

```
Algorithm 1 Extract Twitter sentiment
 1: procedure TWITTER-CONNECTION()
 2:     consumer − key =' xxxxxxxx'
 3:     consumer − secret =' xxxxxxxx'
 4:     access − token =' xxxxxxxx'
 5:     access − token − secret =' xxxxxxxxx'
 6:     self.auth = OAuthHandler(consumer − key, consumer − secret)
 7:     self.auth.set − access − token(access − token, access − token − secret)
 8:     self.api = tweepy.API(self.auth)
 9: end procedure
10:
11: procedure TWEET-CLEANING(t)
12:     tweet = t.remove − Stop − words
13:     Return tweet
14: end procedure
15:
16: procedure TWEET-CLASSIFICATION(t)
17:     t = Tweet − Cleaning(t)
18:     tweet − polarity = t.sentiment.polarity
19:     tweet − polarity
20: end procedure
21:
22: procedure GET-TWEETS(q, count)
23:     fetched − tweets = self.api.search(q = query, count = count)
24:     Return fetched − tweets
25: end procedure
26:
27: procedure MAIN()
28:     st = SentimentalTwitter()
29:     tweets = st.fetch − tweets(query =' politics', count = 300)
30:     PositiveTweets = tweetsthatsentiment =' positive'
31:     NegativeTweets = tweetsthatsentiment =' negative'
32:
33:     for tweet t in PositiveTweets do
34:         print(t)
35:     end for
36:     for tweet t in NegativeTweets do
37:         print(t)
38:     end for
39: end procedure
```

**Figure 1. Extract Sentiment from Twitter**

### A. Implementation

In this paper, we used python to implement sentimental analysis. Some packages have utilized including *tweepy* and *textblob*. We can install the required libraries by following commands:

- pip install tweepy
- pip install textblob

The second step is downloading the dictionary by running the following command:

*python -m textblob.download_corpora.*

The textblob is a python library for text processing and it uses NLTK for natural language processing [6]. Corpora is a large and structured set of texts which we need for analyzing tweets.

### B. Connect to Twitter using APIs

To connect to Twitter and query latest tweets, we need to create an account on twitter and define an application. Users need to go to the apps.twitter.com/app/new and generate the api keys.

The Application seetings is shown in the figure 2. Due to the security reasons the api keys are not shown.

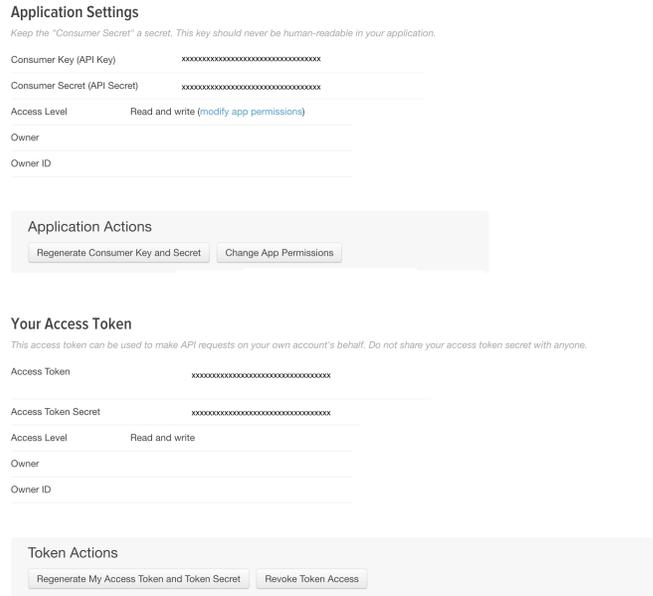

**Figure 2. Twitter Application Management**

### C. Sample Results

Following shows the sample output of the program for the 'fake news' as a query based on the last 300 tweets from Twitter.

Positive tweets percentage: 16.39 %
Negative tweets percentage: 72.13 %
Neutral tweets percentage: 11.47 %

Positive tweets:
tweet: @Nigel_Farage @PoppyLegion Least we forget: Farage is rich. Brexit makes him richer. He is establishment. He is a l… https://t.co/FhZSCBVHJs
tweet: @kirk0071 @Scavino45 @WhiteHouse @POTUS @realDonaldTrump Thanks for the good belly laugh this morning. Your HateTru… https://t.co/AWHXoC84LJ
tweet: @rolandsmartin Roland I like you brother but you really need to distant yourself from Donna Brazile,she's been comp… https://t.co/zqRCsVu98d

Negative tweets:
tweet: RT @Independent: If you saw these tweets, you were targeted by Russian Brexit propaganda https://t.co/Cc8IvQApbY
tweet: Behind Fox News' Baseless Seth Rich Story: The Untold Tale https://t.co/TXcDP1oQ5H

tweet: RT @JackPosobiec: Fake news called the Poland indpendence day parade a "Nazi march." Sick https://t.co/OZA3xUopl1

Table I shows the sentimental analysis results based on different queries including movie, politics, fashion, and fake news. The bar chart, as shown in figure 3, illustrates the data based on the results we got form this step. If we run the program in different times we may get different results, small variance, based on the tweets we fetch. We run the program three times and these results are the average of the outputs.

As it can be clearly seen in the table and diagram the percentage of the neutral tweets are significantly high. This is also important to mention that depends on the data of the experiment we may get different results as people's opinion may change depends on the world circumstances for example fake news as it becomes the world of the year in 2017. For some queries, the neutral tweets are more than 60% which clearly shows the limitation of the current works.

TABLE I.  SENTIMENT ANALYSIS RESULTS

| Query | Positive | negative | Neutral |
|---|---|---|---|
| Movie | 53 | 11.1 | 35.8 |
| politics | 26.6 | 12.2 | 61.1 |
| fashion | 38.8 | 13.3 | 47.7 |
| fake news | 16.3 | 72.1 | 11.4 |
| Justice | 35.2 | 15.9 | 48.8 |
| Humanity | 36.9 | 33.3 | 29.7 |

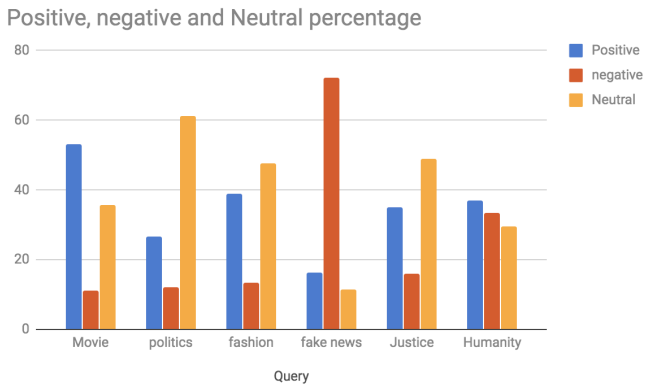

Figure 3. Sentiment results on different queries

## V. CONCLUSION

In this technical paper, we discussed the importance of social newtowk analysis and its applications in different areas. We focused on Twitter as and have implemented the python program to implement sentimental analysis. We showed the results on different daily topics. We realized that the neutral sentments are significantly high which shows there is a need to improve Twitter sentiment analysis.

## REFERENCES

[1] Boguslavsky, I. (2017). Semantic Descriptions for a Text Understanding System. In Computational Linguistics and Intellectual Technologies. Papers from the Annual International Conference "Dialogue"(2017) (pp. 14-28).

[2] Pak, A., & Paroubek, P. (2010, May). Twitter as a corpus for sentiment analysis and opinion mining. In *LREc* (Vol. 10, No. 2010).

[3] Scott, J. (2011). Social network analysis: developments, advances, and prospects. *Social network analysis and mining*, *1*(1), 21-26.

[4] Statista, 2017, https://www.statista.com/statistics/282087/number-of-monthly-active-twitter-users/

[5] Wang, H., Can, D., Kazemzadeh, A., Bar, F., & Narayanan, S. (2012, July). A system for real-time twitter sentiment analysis of 2012 us presidential election cycle. In *Proceedings of the ACL 2012 System Demonstrations* (pp. 115-120). Association for Computational Linguistics.

[6] TextBlob, 2017, https://textblob.readthedocs.io/en/dev/

[7] Pang, B., & Lee, L. (2008). Opinion mining and sentiment analysis. *Foundations and Trends® in Information Retrieval*, *2*(1–2), 1-135.

[8] Dos Santos, C. N., & Gatti, M. (2014, August). Deep Convolutional Neural Networks for Sentiment Analysis of Short Texts

[9] Wilson, T., Wiebe, J., & Hoffmann, P. (2005, October). Recognizing contextual polarity in phrase-level sentiment analysis. In *Proceedings of the conference on human language technology and empirical methods in natural language processing*(pp. 347-354). Association for Computational Linguistics.

[10] Liu, B. (2012). Sentiment analysis and opinion mining. *Synthesis lectures on human language technologies*, *5*(1), 1-167.

[11] Agarwal, A., Xie, B., Vovsha, I., Rambow, O., & Passonneau, R. (2011, June). Sentiment analysis of twitter data. In *Proceedings of the workshop on languages in social media* (pp. 30-38). Association for Computational Linguistics.

[12] Rosenthal, S., Farra, N., & Nakov, P. (2017). SemEval-2017 task 4: Sentiment analysis in Twitter. In *Proceedings of the 11th International Workshop on Semantic Evaluation (SemEval-2017)*(pp. 502-518).

[13] Kiritchenko, S., Zhu, X., & Mohammad, S. M. (2014). Sentiment analysis of short informal texts. *Journal of Artificial Intelligence Research*, *50*, 723-762.

[14] Balahur, A., Steinberger, R., Kabadjov, M., Zavarella, V., Van Der Goot, E., Halkia, M., ... & Belyaeva, J. (2013). Sentiment analysis in the news. *arXiv preprint arXiv:1309.6202*.

[15] Poria, S., Cambria, E., & Gelbukh, A. (2015). Deep convolutional neural network textual features and multiple kernel learning for utterance-level multimodal sentiment analysis. In *Proceedings of the 2015 Conference on Empirical Methods in Natural Language Processing* (pp. 2539-2544).

[16] Ortigosa, A., Martín, J. M., & Carro, R. M. (2014). Sentiment analysis in Facebook and its application to e-learning. *Computers in Human Behavior*, *31*, 527-541.

[17] Bagheri, Hamid, and Abdusalam Abdullah Shaltooki. "Big Data: challenges, opportunities and Cloud based solutions." International Journal of Electrical and Computer Engineering 5.2 (2015): 340.

[18] Bagheri, Hamid, Mohammad Ali Torkamani, and Zhaleh Ghaffari. "Multi-Agent Approach for facing challenges in Ultra-Large Scale systems." International Journal of Electrical and Computer Engineering 4.2 (2014): 151.

[19] Ortigosa, A., Martín, J. M., & Carro, R. M. (2014). Sentiment analysis in Facebook and its application to e-learning. *Computers in Human Behavior*, *31*, 527-541.

[20] Feldman, R. (2013). Techniques and applications for sentiment analysis. *Communications of the ACM*, *56*(4), 82-89.

[21] Dasgupta, S. S., Natarajan, S., Kaipa, K. K., Bhattacherjee, S. K., & Viswanathan, A. (2015, October). Sentiment analysis of Facebook data using Hadoop based open source technologies. In *Data Science and Advanced Analytics (DSAA), 2015. 36678 2015. IEEE International Conference on* (pp. 1-3). IEEE.

[22] Trinh, S., Nguyen, L., Vo, M., & Do, P. (2016). Lexicon-based sentiment analysis of Facebook comments in Vietnamese language. In *Recent developments in intelligent information and database systems* (pp. 263-276). Springer International Publishing.

[23] Haddi, E., Liu, X., & Shi, Y. (2013). The role of text pre-processing in sentiment analysis. *Procedia Computer Science*, *17*, 26-32.